# A Numerical Study of Scaling Issues for Schottky Barrier Carbon Nanotube Transistors

Jing Guo, Supriyo Datta and Mark Lundstrom
School of Electrical and Computer Engineering,
Purdue University, West Lafayette, IN 47907
Email: guoj@purdue.edu

*Abstract*

We performed a comprehensive scaling study of Schottky barrier carbon nanotube transistors using self-consistent, atomistic scale simulations. We restrict our attention to Schottky barrier carbon nanotube FETs whose metal source/drain is attached to an intrinsic carbon nanotube channel. Ambipolar conduction is found to be an important factor that must be carefully considered in device design, especially when the gate oxide is thin. The channel length scaling limit imposed by source-drain tunneling is found to be between 5nm and 10nm, depending on the off-current specification. Using a large diameter tube increases the on-current, but it also increases the leakage current. Our study of gate dielectric scaling shows that the charge on the nanotube can play an important role above threshold.

I.  INTRODUCTION

Carbon nanotube field-effect transistors (CNTFETs) with promising device characteristics have recently been demonstrated [1, 2], so the question of the ultimate device performance capability and minimum device size that might be achievable from an optimized technology becomes important. A recent scaling study of Schottky barrier (SB) CNTFETs by Heinze et al. [3] examined the role of scaling the gate oxide thickness down and the dielectric constant up. That study found that the device performance depends in an unexpected way (as compared to a silicon MOSFET) on the gate oxide thickness and dielectric constant. In a very recent study, the issue of drain voltage scaling has been considered [4]. In this paper, we extend previous work by using a coupled Poisson-quantum transport model to treat the charge in the nanotube self-consistently. A comprehensive study of CNTFET scaling issues is performed to examine the role of gate insulator thickness and dielectric constant, nanotube diameter, Schottky barrier height, drain voltage, and channel length. In contrast to SB CNTFETs with thick gate oxides, SB CNTFETs with thin gate oxides show very strong ambipolar I-V characteristics, even if the barrier heights for electrons and holes are highly asymmetric. The ultimate scaling limit for the channel length imposed by source-drain tunneling is established. The nanotube diameter and drain voltage are shown to have a strong influence on the leakage current. In contrast to a previous study [Hei03] that examined the subthreshold and near threshold regions, we find that increasing the gate dielectric constant improves device performance (the on-current).

In this study, we restrict our attention to Schottky barrier CNTFETs, which operate by modulating the tunneling current at the source contact. (SB CNTFETs are common in experiments at this stage [5].) Note, however, that there are recent reports that CNTFETs without Schottky barriers, which operate more like MOSFETs, can be realized [6]. We assume ballistic transport and solve the Schrödinger equation self-consistently with the Poisson equation. Because our interest is ultimate limits, we assume a coaxial geometry, rather than the planar geometry of the actual devices that have been reported. The coaxial geometry provides the best electrostatic control by the gate and, therefore, the minimum channel length for electrostatic consideration [7, 8]. A zigzag nanotube is assumed, and an atomistic description in terms of $p_z$ orbitals is used.

## II. APPROACH

To investigate the performance of aggressively scaled CNTFETs, we simulated a coaxially gated CNTFET with a 15nm ballistic channel, as shown in Fig. 1. The nominal device has a 2nm $ZrO_2$ gate oxide (a high-K gate insulator of this type has been experimentally demonstrated [1]). The diameter of the (13,0) nanotube is *d≈ 1 nm*, which results in a bandgap of *$E_g$≈ 0.83 eV*. A power supply voltage of 0.4V is assumed, according to the value specified for the 10nm scale MOSFET in ITRS roadmap [9]. The device parameters here are the nominal ones; we explore various issues by varying these parameters.

Carbon nanotube field-effect transistors were simulated by solving the Schrödinger equation using the non-equilibrium Green's function (NEGF) formalism [10] self-consistently with the Poisson equation. Ballistic transport was assumed. An atomistic description of the nanotube using a tight binding Hamiltonian with an atomistic ($p_z$ orbital) basis was used. The atomistic treatment was computationally expensive, but significant computational savings were achieved by the mode space approach [11]. Because the carbon nanotube is coaxially gated, the eigenstates around the tube circumferential direction (modes) are plane waves with wave vectors satisfying the periodic boundary conditions. The two-dimensional nanotube lattice of a (n, 0) zigzag CNT was transformed to n decoupled one-dimensional modes by doing a basis transform from the real space to the mode space in the circumferential direction (essentially Fourier transform). Under typical bias conditions, the few modes that are relevant to electronic transport are treated.

The mode space approach reduces computation significantly yet retains atomistic resolution along the transport direction. For the *i*th mode, the charge density is computed by integrating the local density-of-states (LDOS) over energy,

$$Q_i(z) = (-e)\int_{-\infty}^{+\infty} dE \cdot \text{sgn}[E - E_N(z)]\{D_{iS}(E,z)f(\text{sgn}[E - E_N(z)](E - E_{FS}))$$

$$+ D_{iD}(E,z)f(\text{sgn}[E - E_N(z)](E - E_{FD}))\}, \qquad (1)$$

where $e$ is the electron charge, $\text{sgn}(E)$ is the sign function, $E_{FS,D}$ is the source (drain) Fermi level, and $D_{iS,D}(E,z)$ is the LDOS due to the source (drain) contact as computed by the NEGF method [10]. Because the nanotube conduction and valence bands are symmetric, the charge neutrality level, $E_N(z)$, lies at the middle of band gap [13].

The Schottky barriers at the metal/CNT interfaces were treated phenomenologically. To mimic the continuous states injected from metal to the semiconducing nanotube modes, each semiconducting mode is coupled to the metallic mode of metallic zigzag CNTs at the M/CNT interface with the coupling described by two parameters. The first one is the band discontinuity at the interface, which is the Schottky barrier height when there are no interface states. The second parameter is the tight-binding parameter between the semiconducting and the metallic mode, which determines to the density of metal-induced-gap-states (MIGS). This simple model describes the interface at a similar level as the M/CNT models in literature with the band discontinuity and density of interface states treated as input parameters [14, 15].

A 2D Poisson equation is solved to update the charge neutrality level in eqn (1), $E_N(z) = -e\phi(z, r = d/2)$, where $d$ is the nanotube diameter and $\phi(z)$ is the electrostatic potential,

$$\nabla^2 \phi(z,r) = -\frac{\rho}{\varepsilon}. \tag{2}$$

The potentials at source/drain and gate electrodes are fixed as the boundary conditions, and the gate flat band voltage was assumed to be zero for simplicity. (In practice, it would depend on the gate workfunction.) In order to treat an arbitrary charge distribution on the nanotube channel, the Poisson equation (eqn. (2)) is solved by the method of moments [16]. The iteration between the atomistic quantum transport equation and the electrostatic equation continues until self-consistency is achieved, (a non-linear form of eqn (2) is used to improve the iteration convergence [12]), then the source-drain ballistic current is computed by

$$I = \frac{4e}{h} \int dE \cdot T(E)[f(E - E_{FS}) - f(E - E_{FD})], \tag{3}$$

where $T(E)$ is the source-drain transmission calculated by the NEGF formalism [10]. The gate leakage current is omitted in this study.

## III. RESULTS

We begin by simulating the nominal device and display the resulting log $I_D$ vs. $V_{GS}$ characteristic in Fig. 2a. For the CNTFET with the metal Fermi level at the middle of the bandgap, the transistor is ambipolar, showing symmetric electron and hole conduction (see the solid line in Fig. 2a). The minimum current occurs when the gate voltage is one-half the drain voltage at which the gate-to-source voltage equals the drain-to-gate voltage, and the conduction and valence band profiles are symmetric (see the solid line in Fig. 2b). Radosavljevic et al. observed similar behavior, and this bias is also the optimum bias for observing optical emission in CNTFETs [17]. Several questions will be addressed in the remainder of this paper. What controls the minimum current, the on current, and the subthreshold swing? Can conduction of one type be suppressed so that SB CNTFETs can be used in conventional CMOS digital circuits? How does device performance depend on the nanotube diameter, power supply, gate insulator thickness and dielectric constant, and the channel length?

Figure 2 shows the effect of the metal/CNT barrier height on the $I_D$-$V_{GS}$ characteristics for the nominal device with a thin (2nm) high-K (25) gate dielectric. Reducing the barrier height for electrons to zero increases the electron conduction current for $V_{GS} > V_{DS}/2$ and decreases the hole current for $V_{GS} < V_{DS}/2$. The $I_D - V_{GS}$ characteristic, however, remains approximately symmetrical; the dash-dot line in Fig. 2b explains why. Although the barrier height for holes is high when $\phi_{bn} = 0$ ($\phi_{bp} = E_g$) and barriers to hole conduction exist at both the source and drain electrodes, the barriers are thin. (The thickness of the Schottky barrier is approximately the thickness of the gate oxide [18]). The thin barriers are quite transparent at negative gate voltages. The observation that CNTFETs with thin gate oxide tend to be ambipolar with nearly symmetrical characteristics is consistent with recent experiments [4].

To further clarify the effect of Schottky barrier height on ambipolar conduction, we translated the I-V characteristics of CNTFETs in Fig. 2 along the x-axis so that the minimum current ($I \sim 6 \times 10^{-4} \mu A$) occurred at $V_G = 0V$. (Translating the $I_D$ vs. $V_G$ curve along x-axis in this way could be achieved in practice by adjusting the gate work function). The translated I-V characteristics are shown in Fig. 3. In the subthreshold region, the I-V characteristics of the zero barrier and the mid gap CNTFETs are nearly identical and the minimum leakage current is similar. It is interesting to note that the subthreshold swing is close to the ideal value of 60mV/dec for thermal injection over a barrier, no matter what Schottky barrier height is used. For $V_{GS} = V_{DD}$, the zero barrier height CNTFET delivers more on-current, and for $V_{GS} = -V_{DD}$, it deliver less (hole) on-current that for the mid-gap barrier. The general conclusion, however, is that the results are surprisingly symmetrical about the minimum current – no matter what the barrier height is.

The reason for the near-ideal subthreshold swing can be explained as follows. When the gate oxide is thin, the Schottky barrier is also thin and is essentially transparent to carriers. The current is, therefore, limited by the thermionic emission over a barrier with the height of the barrier determined by the conduction (valence) band in the interior of the channel. Tunneling through the M/S barrier varies with the barrier height and the bias, but it only plays a minor role (because the barrier is so transparent) compared to the barrier in the CNT body. Accordingly, the subthreshold swing is relatively independent of the barrier height, and the best that can be achieved is no better than what could be obtained in a MOSFET. Above the threshold, the situation is different because the barrier between the source and the CNT body is very small, so the tunneling resistance limits the on-current. In this case, the zero barrier contact delivers more on-current.

Things change when the gate insulator is thick. Figure 4 shows the $I_D$ vs. $V_G$ characteristics of a mid-gap SB CNTFET and a $\phi_{bn} = 0$ SB CNTFET with a 40nm-thick K = 25 gate oxide and 100nm channel length. In striking contrast to the thin oxide case, the I-V characteristics of these two CNTFETs are quite different. The minimum leakage current of the zero barrier CNTFET is smaller than for the thin oxide transistor, and the on-off current ratio is much better. The reason is that thicker gate insulators lead to thicker Schottky barriers so the tunneling resistance plays an important role. For the mid-gap CNTFET, the current is always limited by the Schottky barrier at the M/S contact, and the gate fringing field modulates the current by changing the tunneling barrier thickness. As a result, the subthreshold swing is ~200mV/dec – much larger than the theoretical minimum. For the zero barrier height CNTFETs, however, current modulation is achieved by modulating a thermionic barrier inside the CNT body, a mechanism similar to the conventional MOSFET. As the result, the subthreshold swing is much smaller, 90mV/dec. (This value is still larger than the ideal subthreshold swing of 60mV/dec because of short channel electrostatics and the parasitic capacitance between the source/drain contact and the channel that is large [19].) Because the SB is thick, an asymmetric barrier height leads to quite asymmetric electron and hole conduction. For the zero barrier CNTFET, electron conduction is much better than hole conduction. Unfortunately, the thick oxide device displays a rather large subthreshold swing, and the on-current performance suffers from the tunneling barrier.

We turn next to the role of the nanotube diameter in determining the I-V characteristics. Figure 5 shows the $I_D$ vs. $V_G$ characteristics of the CNTFETs with three different nanotube diameters. We assume a mid-gap barrier height for all tubes, which corresponds to the same metal contact material if the work function of an intrinsic tube is independent of the tube diameter. Using a large diameter tube reduces the band gap and significantly increases the minimum leakage current at the ambipolar bias point. At the same time, the on-current is also improved, but the on-off ratio decreases significantly as the nanotube diameter increases. The small band gap of large diameter tube also leads to strong ambipolar conduction even if the gate oxide is thick and barrier heights for electrons and holes are asymmetric [20].

We next examine power supply voltage scaling. Figure 6a shows the $I_D$ vs. $V_G$ characteristics of the nominal SB CNTFET with three different power supply voltages. Note that the minimum current increases exponentially with power supply voltage (as Radosavljevic et al. observed [4]). The reason is that the minimum leakage is achieved when the effective gate to source voltage is one half of the power supply voltage. Reducing the power supply voltage reduces the effective gate to source voltage at the minimum leakage point, thus exponentially reduces the minimal leakage current. Figure 6a also shows that the on-current increases with $V_{DD}$. The off-current vs. on-current for different power supply voltages is plotted in Fig. 6b. The trade-off for reducing the off-current by lowering the power supply voltage is the degradation of on-current. The choice of power supply voltage will depend on the type of circuit applications. For a low power design, the off-current must be small and the on-off current ratio needs to be large, which sets an upper limit for the power supply voltage [4]. Although the low power supply voltage guarantees a small leakage current and large on-off ratio (because for small power supply voltage, the transistor operates in the exponential portion of the $I_D$ vs $V_G$ characteristics), the on-current may still be too small for high-performance applications. To achieve a large on-current and a reasonable off-current simultaneously, which is required for high performance applications, the power supply voltage must be large enough.

Figure 7 explores the issue of channel length scaling. In order to establish the ultimate scaling limit imposed by source-drain tunneling, very thin gate oxide ($t_{ox} = 2nm$ for the channel length of 10nm and longer and $t_{ox} = 1nm$ for the channel length of 5nm) is used to ensure excellent gate controlled electrostatics. Although we assume a mid-gap Schottky barrier height, similar observations apply to other barrier heights. When the channel length is larger than about 15nm, the I-V characteristics are independent of the channel length because the channel is ballistic and the quantum tunneling resistance of the Schottky barrier at the source dominates the total channel resistance. Scaling the channel length down to 10nm significantly increases the off-current, but the on-off current ratio still exceeds 100, which is probably acceptable for digital logic. If the channel length is aggressively scaled down to 5nm, the on-off current ratio decrease to less than 10 due to significant source-drain quantum tunneling. Compared to Si MOSFETs with the similar channel lengths, the tunneling leakage of the CNTFET is more severe, partly due to typically smaller band gap and lighter effective mass for carriers in the CNT channel. (A parabolic E-k fit the very bottom of the conduction band of a 1nm diameter CNT gives an effective mass of ~0.08.)

Figure 8 explores the role of the gate dielectric constant. A previous study, which omitted the charge on the nanotube, concluded that simply increasing the gate dielectric constant everywhere didn't change the band profile thus had no effect on the I-V characteristics of SB-CNTFETs [3]. Those conclusions apply below and near threshold. Figure 8, which shows the $I_D$ vs. $V_G$ characteristics for mid-gap SB-CNTFETs with a 2nm thick oxide and three different dielectric constants, shows that a high-$\kappa$ gate insulator does increase the on-current. The inset, which plots the conduction band profiles for three transistors at the on-state, sheds light on these results. Because the Schottky barriers are thin (due to thin gate oxide), the charge density inside the tube is high. When the gate dielectric constant is low, this charge produces a considerable self-consistent potential. The conduction band floats up, which makes the Schottky barrier thicker and the conduction band in the interior of the channel higher. Because the tunneling current exponentially depends on the Schottky barrier thickness, the on-current of transistors with low gate dielectric constants is smaller. Calculations which omit the charge on the CNT overestimate the current when the mobile charge is important.

## IV. DISCUSSION

The key point of this work, as also pointed out by Radosavljevic et al. [4], is the central importance of ambipolar conduction in SB CNTFETs when the gate oxide is thin, as it must be for high-performance transistors. To use such SB CNTFETs in conventional CMOS circuits, will require careful device design because negative gate to source voltages, and, therefore, high leakage currents, would result when transistors are stacked. The results presented in Sec. III shed light on how the leakage and on-current varies with device parameters such as gate insulator thickness, nanotube diameter, power supply voltage, etc. It is possible that gate work function engineering could be employed so that in the bias region of interest, only one branch of the ambipolar I-V is used, but this will be difficult because $V_{GS} < 0$ occurs for stacked transistors. Alternatively, one could explore MOSFET-like devices for which ambipolar conduction would not occur.

Finally, we should note that we also examined gate oxide thickness scaling and the effect of contact geometry. The results were similar to those of previous studies [3, 5] (i.e. reducing the gate oxide thickness and contact size improves the subthreshold swing) were observed.

## IV. CONCLUSIONS

In summary, scaling issues for SB-CNTFETs were explored by self-consistent, atomistic scale simulations. Ambipolar conduction was found to be an important feature that must be carefully considered in transistor design. The minimum subthreshold swing is 60 mV/dec., just as it is for a MOSFET. The scaling limit for CNTFETs imposed by source-drain tunneling is between 5nm and 10nm and is determined by the small band gap and strong wave behavior of carriers in CNTs. CNTFETs, therefore, offer no scaling advantage over a MOSFET. Larger tube diameter lowers the Schottky barrier height and delivers more on-current, but larger diameter nanotubes also result in larger leakage currents. Using a high-$\kappa$ gate oxide improves the on-current, just as it improves the on-current of a MOSFET. The understanding of CNTFET device physics should prove useful in optimizing device designs.

## ACKNOWLEDGEMENT

It is our pleasure to thank A. Javey and Prof. H. Dai of Stanford for cordially sharing their insights and unpublished experiments with us, which significantly deepen our understanding of CNTFETs. We thank Dr. D. Kienle of Purdue, Dr. M. P. Anantram of NASA Ames, Drs. Ph. Avouris and J. Appenzeller of IBM for extensive technical discussions. This work was supported by the National Science Foundation, grant no. EEC-0085516, the NSF Network for Computational Nanotechnology, and the MARCO Focused Research Center on Materials, Structure, and Devices, which is funded at MIT, in part by MARCO under contract 2001-MT-887, and DARPA under grant MDA972-01-1-0035.

**FIGURES**

Fig. 1: The modeled CNTFET with a coaxial gate. The gate length is the same as the source to drain spacing; the nominal value is 15nm. A 2nm-thick $ZrO_2$ gate insulator and a (13,0) zigzag nanotube (with the diameter $d\sim 1nm$ and the band gap $E_g\sim 0.8eV$) are assumed.

Fig. 2 Transistor characteristics when the gate oxide is thin. (a) $I_D$ vs. $V_G$ characteristics for the nominal CNTFET (as shown in Fig. 1) with three different barrier heights. The Schottky barrier height for electrons is $\phi_{bn} = 0$ for the dash-dot line, $\phi_{bn} = E_g/2$ for the solid line and $\phi_{bn} = E_g$ for the dashed line. (b) The conduction and valence band profile at $V_G = 0.2V$. The dash-dot line is for $\phi_{bn} = 0$ and the solid line is for $\phi_{bn} = E_g/2$. The flat band voltage of all transistors is zero.

Fig. 3 Shifted $I_D$ vs. $V_G$ characteristics for the nominal CNTFET (as shown in Fig. 1) with the barrier height for electrons $\phi_{bn} = 0$ (the solid-dash lines) and $\phi_{bn} = E_g/2$ (the solid lines). The minimal leakage current is shifted to $V_G = 0$ by adjusting the flat band voltage for each transistor. The left axis shows the I-V on log scale and the right axis shows the same curves on linear scale.

Fig. 4  $I_D$ vs. $V_G$ for thick gate oxide (the oxide thickness $t_{ox} = 40nm$ and dielectric constant $\varepsilon = 25$). The channel length is 100nm. The Schottky barrier height for electrons is $\phi_{bn} = 0$ (the solid-dash lines) and $\phi_{bn} = E_g/2$ (the solid lines). A (25,0) nanotube (with a diameter $d$~2.0nm and $E_g$ ~ $0.43eV$) is used as channel.

Fig. 5  *Scaling of nanotube diameter*. $I_D$ vs. $V_G$ characteristics at $V_D$=0.4V for the nominal CNTFET with different nanotube diameter. The solid line with circles is for (13,0) CNT (with d~1nm), the sold line is for (17,0) CNT (with d~1.3nm), and the dashed line is for (25,0) CNT (with d~2nm). The flat band voltage is zero and the Fermi level lies in the middle of the band gap for all transistors.

Fig. 6 *Scaling of Power supply voltage.* (a) $I_D$ vs $V_G$ characteristics under different power supply voltages for the nominal CNTFET (Fig. 1) with mid-gap Schottky barriers. For each power supply voltage, the minimal leakage point is shifted to $V_G = 0$ by adjusting the flat band voltage. The on-current is defined at $V_G = V_D = V_{DD}$. (b) The off-current vs. on-current for different power supply voltages.

Fig. 7 *Channel length scaling.* $I_D$ vs. $V_G$ characteristics of CNTFETs with different channel length. The circles are for channel length $L_{ch} = 30nm$ and gate ZrO$_2$ thickness $t_{ox} = 2nm$, the dash-dot line for $L_{ch} = 15nm$ and $t_{ox} = 2nm$, the solid line for $L_{ch} = 10nm$ and $t_{ox} = 2nm$, and the dashed line for $L_{ch} = 5nm$ and $t_{ox} = 1nm$. The flat band voltage is zero and mid-gap Schottky barriers are assumed for all transistors.

Fig. 8 *Gate dielectric scaling*. $I_D$ vs. $V_G$ characteristics at $V_D$=0.4V for the nominal CNTFET with different gate dielectric constant. The solid line is for $\varepsilon_{ox} = 25$, the dashed line for $\varepsilon_{ox} = 4$ and the dash-dot line for $\varepsilon_{ox} = 1$. The inset shows the corresponding conduction band profile at $V_G$=0.6V. The flat band voltage is zero and mid-gap Schottky barriers are assumed for all transistors.

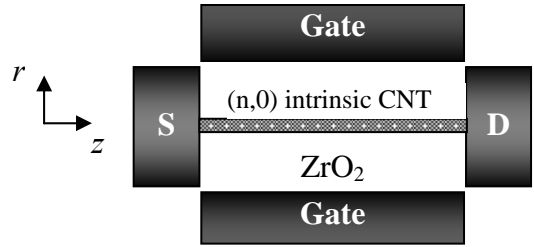

Fig. 1: The modeled CNTFET with a coaxial gate. The gate length is the same as the source to drain spacing; the nominal value is 15nm. A 2nm-thick $ZrO_2$ gate insulator and a (13,0) zigzag nanotube (with the diameter $d \sim 1nm$ and the band gap $E_g \sim 0.8eV$) are assumed.

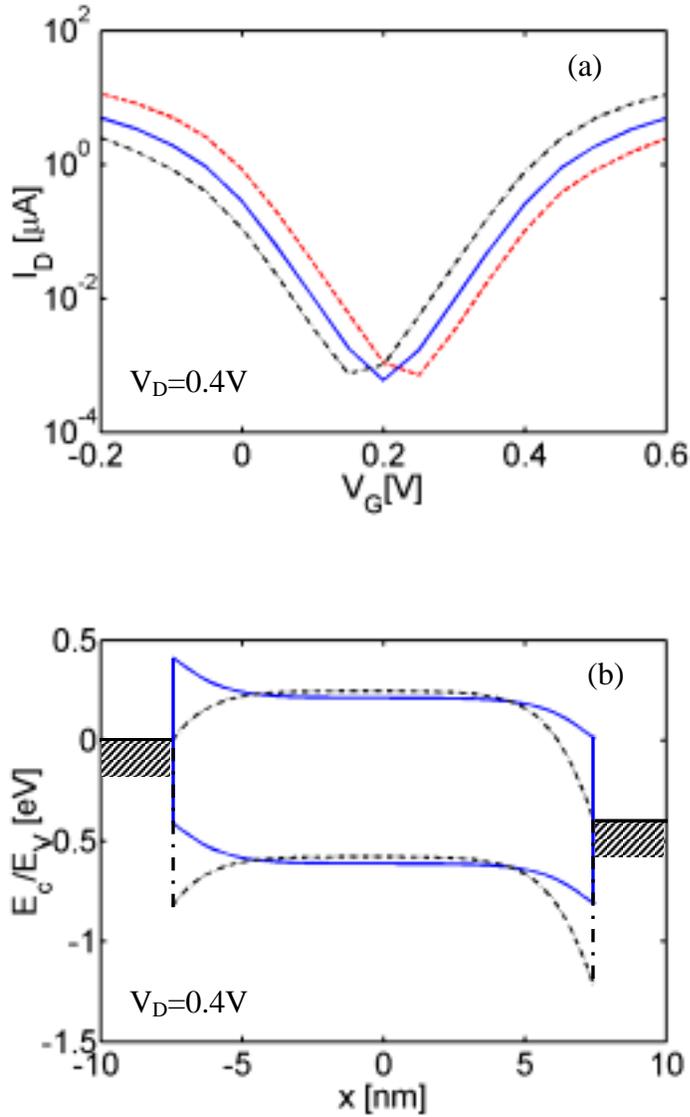

Fig. 2 Transistor characteristics when the gate oxide is thin. (a) $I_D$ vs. $V_G$ characteristics for the nominal CNTFET (as shown in Fig. 1) with three different barrier heights. The Schottky barrier height for electrons is $\phi_{bn} = 0$ for the dash-dot line, $\phi_{bn} = E_g/2$ for the solid line and $\phi_{bn} = E_g$ for the dashed line. (b) The conduction and valence band profile at $V_G = 0.2V$. The dash-dot line is for $\phi_{bn} = 0$ and the solid line is for $\phi_{bn} = E_g/2$. The flat band voltage of all transistors is zero.

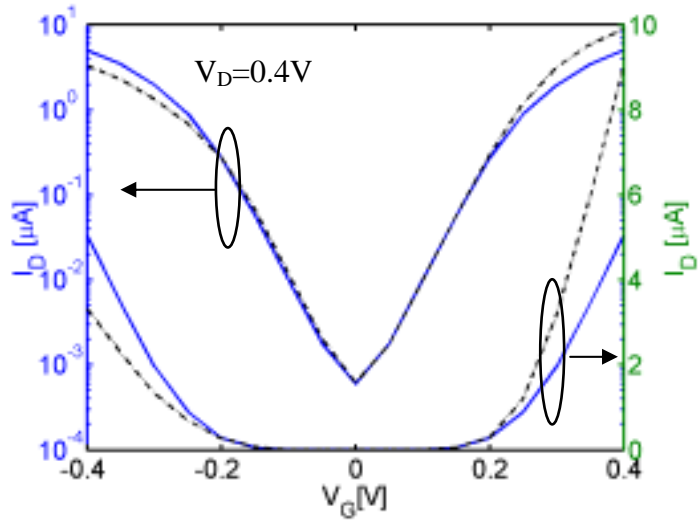

Fig. 3 Shifted $I_D$ vs. $V_G$ characteristics for the nominal CNTFET (as shown in Fig. 1) with the barrier height for electrons $\phi_{bn} = 0$ (the solid-dash lines) and $\phi_{bn} = E_g/2$ (the solid lines). The minimal leakage current is shifted to $V_G = 0$ by adjusting the flat band voltage for each transistor. The left axis shows the I-V on log scale and the right axis shows the same curves on linear scale.

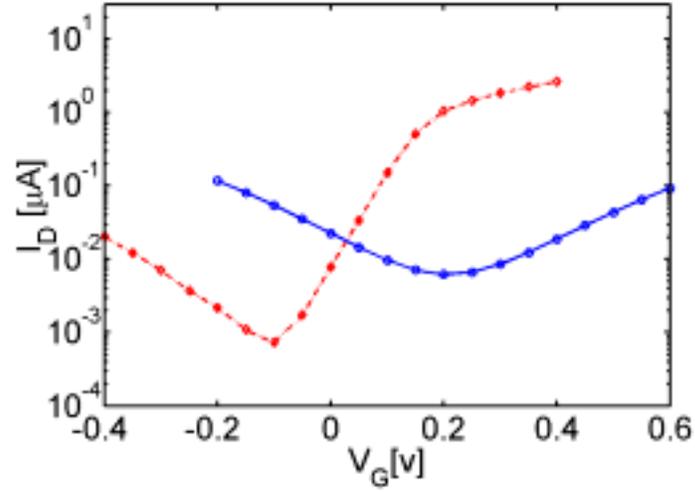

Fig. 4  $I_D$ vs. $V_G$ for thick gate oxide (the oxide thickness $t_{ox} = 40nm$ and dielectric constant $\varepsilon = 25$). The channel length is 100nm. The Schottky barrier height for electrons is $\phi_{bn} = 0$ (the solid-dash lines) and $\phi_{bn} = E_g/2$ (the solid lines). A (25,0) nanotube (with a diameter $d$~2.0nm and $E_g$ ~ $0.43eV$) is used as channel.

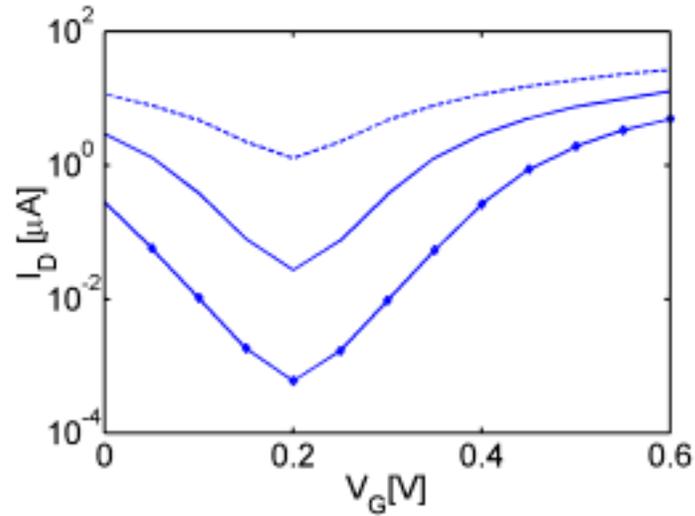

Fig. 5 *Scaling of nanotube diameter*. $I_D$ vs. $V_G$ characteristics at $V_D=0.4V$ for the nominal CNTFET with different nanotube diameter. The solid line with circles is for (13,0) CNT (with d~1nm), the sold line is for (17,0) CNT (with d~1.3nm), and the dashed line is for (25,0) CNT (with d~2nm). The flat band voltage is zero and the Fermi level lies in the middle of the band gap for all transistors.

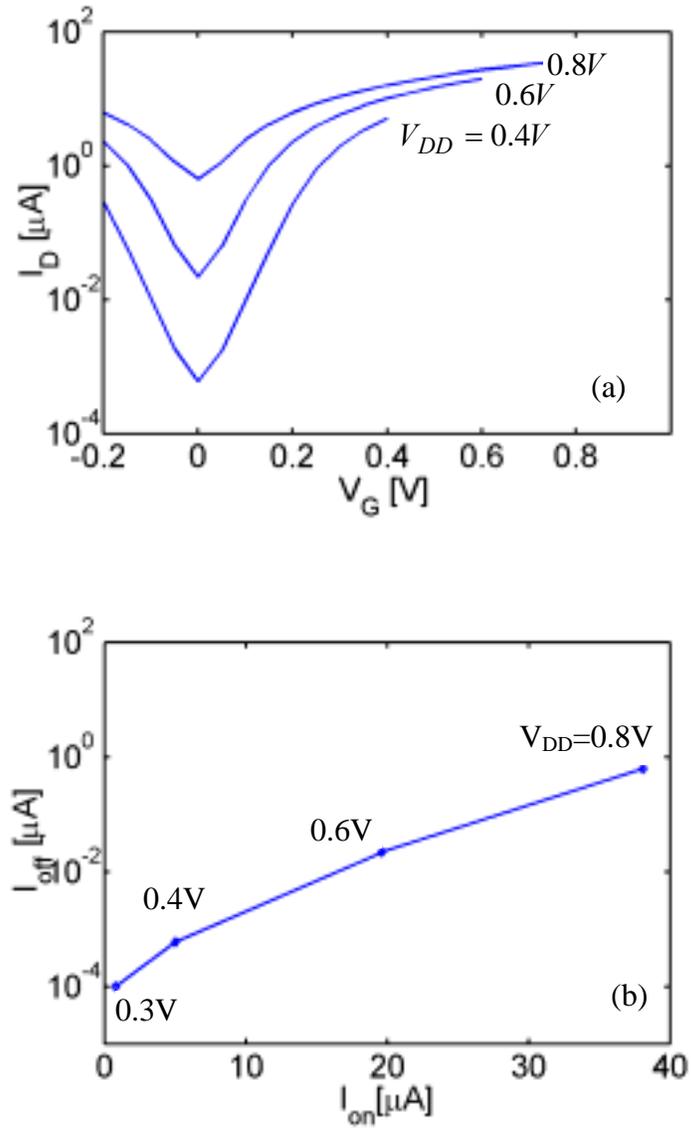

Fig. 6 *Scaling of Power supply voltage.* (a) $I_D$ vs $V_G$ characteristics under different power supply voltages for the nominal CNTFET (Fig. 1) with mid-gap Schottky barriers. For each power supply voltage, the minimal leakage point is shifted to $V_G = 0$ by adjusting the flat band voltage. The on-current is defined at $V_G = V_D = V_{DD}$. (b) The off-current vs. on-current for different power supply voltages.

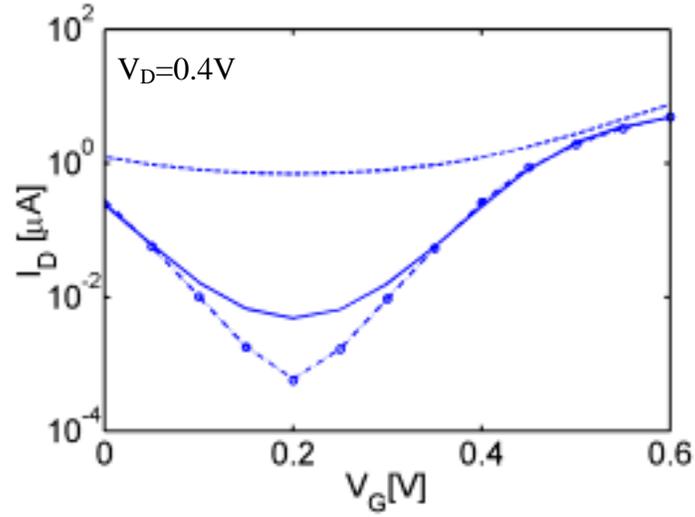

Fig. 7 *Channel length scaling.* $I_D$ vs. $V_G$ characteristics of CNTFETs with different channel length. The circles are for channel length $L_{ch} = 30nm$ and gate $ZrO_2$ thickness $t_{ox} = 2nm$, the dash-dot line for $L_{ch} = 15nm$ and $t_{ox} = 2nm$, the solid line for $L_{ch} = 10nm$ and $t_{ox} = 2nm$, and the dashed line for $L_{ch} = 5nm$ and $t_{ox} = 1nm$. The flat band voltage is zero and mid-gap Schottky barriers are assumed for all transistors.

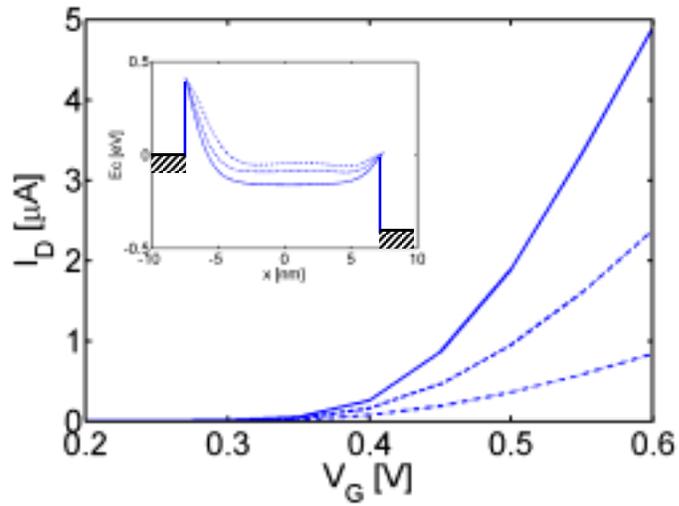

Fig. 8 *Gate dielectric scaling*. $I_D$ vs. $V_G$ characteristics at $V_D$=0.4V for the nominal CNTFET with different gate dielectric constant. The solid line is for $\varepsilon_{ox} = 25$, the dashed line for $\varepsilon_{ox} = 4$ and the dash-dot line for $\varepsilon_{ox} = 1$. The inset shows the corresponding conduction band profile at $V_G$=0.6V. The flat band voltage is zero and mid-gap Schottky barriers are assumed for all transistors.